\documentclass[prl,aps,twocolumn,floatfix]{revtex4}
\usepackage{graphicx}

\def\be{\begin{equation}}
\def\ee{\end{equation}}
\def\ba{\begin{eqnarray}}
\def\ea{\end{eqnarray}}

\def\p{{\bf p}}
\def\q{{\bf q}}

\def\x{{\bf x}}
\def\sf{\sqrt{f}}
\def\D{\Delta}
\def\Rini{R_1}
\def\Rfin{R_2}
\def\Drho{\Delta_{,\rho}}
\def\eps{\epsilon}
\def\emin{\epsilon_{\min}}
\begin{document}
\title{Relaxation dynamics in a strongly coupled Fermi superfluid}
\author{Sergei Khlebnikov}
\email{skhleb@purdue.edu}
\affiliation{Department of Physics and Astronomy, Purdue University, West Lafayette, IN 47907}
%\date{}
\begin{abstract}
The key feature of time-dependent dynamics in a paired Fermi superfluid
is the presence of a large number of independent degrees of freedom---the pairing
amplitudes of fermions with different momenta. We argue that useful 
prototypes of this
dynamics come from D-brane constructions of string theory. 
Using a specific example of that kind, we identify
the mechanism by which a strongly coupled Fermi superfluid relaxes to equilibrium; 
it involves a wave of excitation in the momentum space, propagating
from the Fermi surface towards the ultraviolet. For a sudden quench 
induced by a change in the fermion coupling, we find that the relaxation occurs rapidly, 
over only a few oscillations of the quasiparticle gap. 
\end{abstract}
\maketitle
{\em Introduction.} 
Non-equilibrium, time-dependent behavior of Fermi superfluids (systems where superfluidity is
due to fermion pairing) has recently become
a subject of intensive experimental \cite{Zwierlein&al} and theoretical 
\cite{Barankov&al,Warner&Leggett,Yuzbashyan&al_PRB,Yuzbashyan&al} interest. This is 
due in large part
to the possibilities offered by cold Fermi gases, where the interaction strength can
be modulated in time, either periodically or in a rapid (quenched) fashion. Our aim here
is to identify the principal mechanism by which relaxation occurs in these
systems, in the case when the interaction between the atoms is strong but the coupling
to an external heat bath is weak. 

We begin by recalling that, already in the conventional
mean-field theory, the problem requires introduction of a large 
(in the thermodynamic limit, infinite) number of local fields. These are the pairing 
amplitudes $\psi_\p(\x)$---the Fourier transforms with respect to $\q$ of the
pair expectation values $\langle \eps^{\sigma\sigma'} c_{\p+\q, \sigma} c_{-\p,\sigma'} \rangle$. 
Here $c_{\p\sigma}$ is the annihilation operator
of the fermion with momentum $\p$ and spin projection $\sigma$, and $\eps^{\sigma\sigma'}$ is 
the antisymmetric unit tensor (for definiteness, we assume that the pairing channel is a 
spin singlet). Note that the number of degrees of freedom (DOF) 
remains large---one for each value of $\p$---even for uniform
states, when all $\psi_\p$ are independent of $\x$.

As recognized in earlier theoretical studies 
\cite{Warner&Leggett,Yuzbashyan&al_PRB,Yuzbashyan&al},
this abundance of DOF can lead, under certain
conditions, to an ``ergodic'' behavior: the energy of a non-equilibrium initial state
is redistributed among the DOF, and
the quasiparticle gap approaches a steady value even in the absence of any coupling to
an external heat bath. The steady value found in these studies,
however, is not the equilibrium one and,
in addition, the relaxation to it is extremely slow, requiring
many oscillations of the gap parameter even when the interaction is not weak. 
For instance, for a sudden quench of the interaction strength in a paired ground state, 
Ref.~\cite{Yuzbashyan&al} finds only a $1/\sqrt{t}$ decrease in the amplitude of 
the oscillations with time. In contrast, in the experiment \cite{Zwierlein&al}, the
relaxation is found to be quite rapid: the observed values of the relaxation time correspond 
to at most a few oscillations of the gap.

One may suspect that the reason why the dynamics found in 
\cite{Warner&Leggett,Yuzbashyan&al_PRB,Yuzbashyan&al} is so slow is that the computations
have been based on the reduced BCS
Hamiltonian, in which the pairing potential is taken to be momentum-independent,
$H_{red} = - g \sum_{\p\p'} \psi^\dagger_{\p'} \psi_\p$. In Anderson's pseudospin representation
\cite{Anderson}, this corresponds to different
pseudospins $\psi_\p$ communicating with one another only via coupling to a single
effective field, $\Delta^\dagger \propto \sum_{\p'} \psi^\dagger_{\p'}$. One can imagine that
this single field acts as a bottleneck, preventing the system from achieving the full
many-body behavior.

The assumption of a momentum-independent pairing potential can be justified 
for a weakly coupled superfluid, where the scattering length $|a|$ is much shorter than the 
typical de Broglie wavelength. It necessarily breaks down at strong coupling, when
$|a|$ becomes large, and the amplitude of two-particle scattering depends significantly on the
transferred momentum. We would like to understand if there are any novel effects in the
relaxation dynamics of a superfluid that occur because of that dependence.

We propose that the language in which this question is
most readily discussed is not that of the conventional field theory
but that using D-brane constructions of string theory. Open strings can end on
D branes \cite{Dai:1989ua}, and the ground states of these strings can represent the fermions
in a superfluid. As these strings
interact via gauge fields that live on the D-brane worldvolumes, they
acquire a momentum-depending pairing potential---precisely
the case we need.

It is by now well known that string theory offers novel perspectives on physics of 
strongly coupled quantum systems. One is
holography, by which we mean here the idea that an energy scale
in a quantum theory can be represented as an extra spatial dimension 
\cite{Maldacena:1997re}. In our case, the relevant energy scale is 
the distance between the fermion momentum $\p$ and the Fermi surface. 
By converting this distance into a coordinate, holography makes
the dynamics in the momentum space particularly
easy to visualize. 

Our approach is different from the earlier holographic approach 
\cite{Gubser:2008px,Hartnoll:2008vx}, 
in which a superfluid is obtained as the dual to a black hole with charged scalar
``hair.'' That description
does not capture the proliferation of the DOF noted earlier and therefore will not 
contain any relaxation mechanism that relies on it (other mechanisms, such as 
relaxation via a
coupling to a heat bath, may still be accessible in that approach \cite{Bhaseen:2012gg}). 
Rather, our method has a parallel
with the holographic approach to meson spectroscopy 
\cite{Kruczenski:2003be}, in that we view fluctuations of the pairing
amplitude (``collective'' excitations in Anderson's 
\cite{Anderson} terminology) as vibrations of a D brane.
Moreover, similarly to that case, in the strongly coupled limit, 
the low-energy spectrum near the gapped ground state is due entirely to these 
collective excitations, with individual quasiparticles pushed out to the high-energy part.
(Recall, by way of contrast, that, for the reduced BCS Hamiltonian with a
momentum-independent potential,
small oscillations near the ground state are all
of the ``individual particle'' type, with ``collective'' modes absent altogether 
\cite{Anderson}.)
Unlike in the meson
case, however, the spectrum of the collective
excitations here is continuous rather than discrete. In the present context, this 
means that our superfluid is on the BCS side of the BCS-BEC crossover. 

Let us stress that our goal here is not to present a detailed description 
of a specific atomic superfluid, but rather produce a prototype model that highlights the
possibility of a new type of relaxation dynamics. In particular, it is not clear
if any experimentally 
accessible system exhibits (as our model does) the aforementioned hierarchy of 
energy scales---the one between the
collective and single-particle excitations. 
We expect, however, that our main results are quite generic, namely, that,
even in cases when pair-breaking
effects are important, they will be in addition to the pair-preserving mechanism
identified here.

{\em String theory-superconductivity dictionary.} 
The following dictionary is motivated by a specific example 
of a stringy superfluid---a system of the D3 and D5 branes intersecting over a line
\cite{Khlebnikov:2012yd,Khlebnikov:2013sha}. To be able to present the key features of
the construction more concisely, we have abstracted them
into dictionary entries. For concrete calculations, we will revert to the D3/D5 case.

(i) Fermionic quasiparticles in a superconductor correspond to strings connecting
two different D-branes, one D-brane having $s$ and the other $s'$ spatial dimensions. 
A finite distance between the branes implies a nonzero minimum 
energy $\emin$
for a connecting string; that minimum energy is the quasiparticle gap. Thus, fully gapped
states correspond to branes passing at a finite distance from each 
other, while states that have a gapless (normal) component to branes that intersect.

(ii) A pair of $s/s'$ strings with opposite orientations corresponds to fermions at opposite
points on the Fermi surface (FS). Such a pair can disappear into vacuum, with the
energy taken up by vibrations of the branes.
These vibrations correspond to fluctuations of the pairing amplitude.

(iii) The 10 dimensions of superstring theory split into three groups. Those in
which both D$s$ and D$s'$ branes extend are the dimensions in which quasiparticles can
propagate freely. These are the usual spacetime dimensions, 
their number determined by 
the geometry of the sample.  Dimensions orthogonal to 
both branes are components of the superconducting order parameter. Finally, the radial
distance in the directions
in which one brane extends and the other does not is the holographic or ``energy''
dimension \cite{Maldacena:1997re}.
In our case, it corresponds to the distance between the fermion
momentum and the FS.

{\em The D3/D5 system.} The entries of the dictionary
can be discussed more concretely using as an example the system of D3 and
D5 branes sharing one spatial direction \cite{Khlebnikov:2012yd,Khlebnikov:2013sha}
in the type IIB string theory. We call that direction longitudinal.
The spectrum of 3/5 strings is non-supersymmetric, and the perturbative ground 
state of such a string is a fermion. 
With periodic boundary conditions
in the longitudinal direction, this system describes 
a Fermi superfluid ring. The ring can be made ``thick''
by increasing the number of coincident D3 branes to $N \gg 1$ and interpreting $N$ 
as the number of transverse channels. The basis for this interpretation is that the
large $N$ suppresses quantum fluctuations and enhances superconductivity in the same way
as the cross-sectional area of a thick sample would.

In the limit when the string coupling $g_s$
is small but the 't Hooft coupling $\lambda = g_s N$ is large, the D3 branes can be 
replaced as usual \cite{Polchinski:1995mt} by the classical black brane solution
\cite{Horowitz:1991cd} with $N$ units of the D-brane charge, while
the influence of the D5 on the geometry can be neglected. 
Here we limit ourselves to the case of zero temperature, when the solution is extremal.
Note that the condition $g_s \ll 1$ implies that oscillations of the D5 will not heat up 
the bulk geometry rapidly: on the
time scale of interest to us the extremal solution will remain extremal. 

The density of states at the FS in the (unstable) normal state is
$N L_x / \pi \hbar$, where $L_x$ is the length of the ring;
no further parameters are needed to keep this nonzero \cite{f1}.
Accordingly, unlike other constructions of Fermi surfaces from strings
(see, e.g., \cite{DeWolfe:2011aa}), we do not supply the black brane 
with any charges besides the D-brane charge.

Both purely superconducting (SC) and mixed SC/normal states can be considered in 
this approach, either with or without electric current 
\cite{Khlebnikov:2012yd,Khlebnikov:2013sha}.
Here, motivated primarily by existing experiments with
cold atoms, we consider purely SC states without current. In addition, we assume the
state to be uniform. This means that both the
initial state and the perturbation are independent of the longitudinal coordinate,
the case for instance for a uniform quench. 

The states in question are described by D5 embeddings of the following form. Let the
D3 span the directions
$x^0, x^1, x^2,$ and $x^3$, and the D5 the directions $x^0, x^1, x^4, \ldots, x^7$. The 
directions $x^8$ and $x^9$ are transverse to both branes. 
Define polar coordinates in these directions, $x^8 = \D \cos\phi$,
$x^9 = \D \sin\phi$, and the radial distance $\rho = [(x^4)^2 + \ldots + (x^7)^2]^{1/2}$
in the directions transverse to the D3s but not to the D5. The complex position 
of the D5, $x^8 + i x^9$, is the SC order parameter. Consider solutions in which it
is purely real and depends only on $\rho$ and possibly time, i.e., has the form
$\D = \D(t, \rho)$ and $\phi = 0$. To complete the definition of the embedding, we need
to specify also the form of the D5 worldvolume gauge field $A$. We set
$A= 0$ as appropriate for ground states without current \cite{Khlebnikov:2012yd}.

Consider a radial string connecting the D3s to some static D5 profile $\D(\rho)$.
The classical energy of such a string is
\be
\eps(\rho) = \frac{1}{2\pi \alpha'} [\rho^2 + \D^2(\rho)]^{1/2} \, ,
\label{eps}
\ee
where $1/2\pi \alpha'$ is the string tension. Quantum corrections to (\ref{eps})
are small under the conditions relevant here.
We see that $\rho / 2\pi \alpha'$ plays the role of 
$v_F |p - p_F|$, the distance between the quasiparticle
momentum and the FS.
By definition, $\rho \geq 0$; this range
describes both the states above the Fermi surface and
those below, as the fermionic ground state of 
3/5 strings is two-component. 

The action for a D5 embedded into the D3 geometry, under the conditions specified above,
is given by the Dirac-Born-Infeld term \cite{Leigh:1989jq}, which takes the form
\be
S_{D5} = - 2\pi^2 \tau_5 L_x \int dt  d\rho \rho^3  \sf 
\left( 1 - f \D_{,t}^2 + \Drho^2 \right)^{1/2} \, .
\label{SD5}
\ee
Subscript commas denote partial derivatives, 
$\tau_5$ is the D5 tension, and $f = 1 + R^4/r^4$, with $r = (\rho^2 + \D^2)^{1/2}$ and
\be
R^4 = 4 \pi g_s N (\alpha')^2 \, ,
\label{R4}
\ee
is the metric function describing the D3 background. 
The near-horizon limit, in which the system becomes dual
to a field theory \cite{Maldacena:1997re}, is $r \ll R$ but is not taken 
here. Indeed, the part of the D5 at $r > R$ is interpreted as describing
the pairing amplitudes far from the Fermi surface, and we will find that
these pairing amplitudes
play an important role in relaxation of the order parameter. 

The stable equilibrium profile of the D5, $\D^{(eq)}(\rho; R)$, is nontrivial 
and corresponds
to a superconducting ground state \cite{Khlebnikov:2012yd}. It has a typical magnitude
$\Delta \sim R$, varies on the scale $\rho \sim R$, and decays to zero at 
$\rho\to \infty$. 
The quasiparticle gap is given by the minimum 
of (\ref{eps}) with $\D = \D^{(eq)}$, with respect to $\rho$. It is finite and corresponds to
$\rho = 0$. 

{\em Boundary conditions} (BCs).
At $\rho = 0$, regularity of $\D$ requires $\Drho(t,0) = 0$. At large $\rho$, to mimic 
the effect of the asymptotic region $\rho\to \infty$ in a finite-size calculation, we impose 
the absorptive BC
\be
\Drho = - \D_{,t} - \frac{2 \D}{\rho} \, .
\label{bc}
\ee
For time-dependent solutions, the first term on the right-hand side is the dominant one,
and (\ref{bc}) selects the outgoing wave $\D \sim \chi(t - \rho)$. For the equilibrium
state, only the second term remains; 
it produces the correct asymptotics $\D^{(eq)} \sim 1 / \rho^2$.
 
{\em Uniform quench.} Suppose that at $t=0$ the D5 is in the equilibrium
state corresponding to the D3 background with a certain $R=\Rini$, that is
$\D(0,\rho) = \D^{(eq)}(\rho; \Rini)$ and $\D_{,t}(0,\rho) = 0$. Meanwhile, 
the ``true'' value of the background
metric parameter is some other $R=\Rfin$. The deviation of $\D(0,\rho)$ from the true
equilibrium will trigger classical evolution governed by (\ref{SD5}). 
According to (\ref{R4}), this evolution can be viewed as a result of 
an abrupt change in the string coupling $g_s$; that corresponds to a change in 
the interaction strength in experiments with cold atoms. 

The large value
of $\tau_5$ suppresses quantum fluctuations of the D5's shape. As long as deviations
from equilibrium remain larger than these fluctuations, the classical approximation
is sufficient. On the other hand, unless the initial deviation is too large, the D5 never
crosses the D3s, and the 3/5 strings remain too heavy to be efficiently produced by
the oscillating D5: the single-particle excitations decouple.

\begin{figure}
\begin{center}
\includegraphics[width=3.4in]{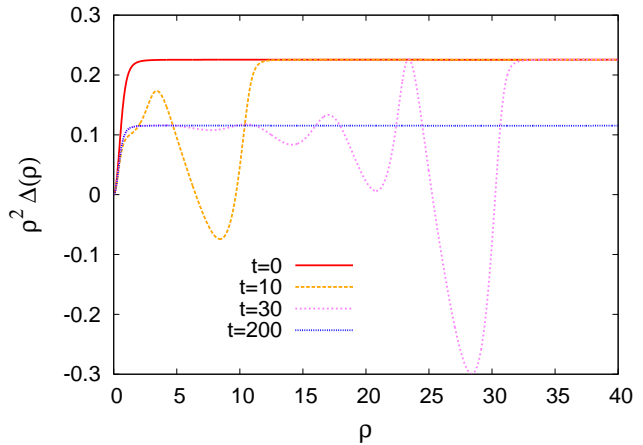}
\end{center}                                              
\caption{The relaxation mechanism: a wave of excitation
propagating along the D5 to large values of $\rho$. Since $\rho$ maps to the distance
from the Fermi surface, in the dual superfluid, this is a wave in the momentum space.
}                                              
\label{fig:wave}                                                                       
\end{figure}

In what follows, we measure all lengths and times in units of $\Rini$, i.e., set $\Rini =1$.
Numerically computed evolution of the order 
parameter for $\Rfin = 0.8$ is shown
in Fig.~\ref{fig:wave}. To emphasize the role of large $\rho$,
we plot $\rho^2 \D$ rather than $\D$ itself. 
We see that the order parameter relaxes to the true equilibrium profile (the lower
mostly horizontal line in Fig.~\ref{fig:wave}). This
occurs via propagation of a wave of excitation towards large $\rho$.
Recall that $\rho$ corresponds to the separation
of the fermion momentum from the FS. We conclude that the relaxation process
activates the pairing
amplitudes of fermions far from the FS. Although small in the equilibrium state,
these amplitudes are essential for the transfer
of energy progressively further away from the FS, in a process not unlike
a (decaying) turbulent cascade \cite{f3}.

Significantly, 
Fig.~\ref{fig:wave} reveals no stable breathing mode at small $\rho$. Such a mode would
correspond to a bound state of two fermions. In application
to cold atoms, its absence means that our system corresponds to 
the BCS side of the BCS-BEC crossover. 

\begin{figure}
\begin{center}
\includegraphics[width=3.4in]{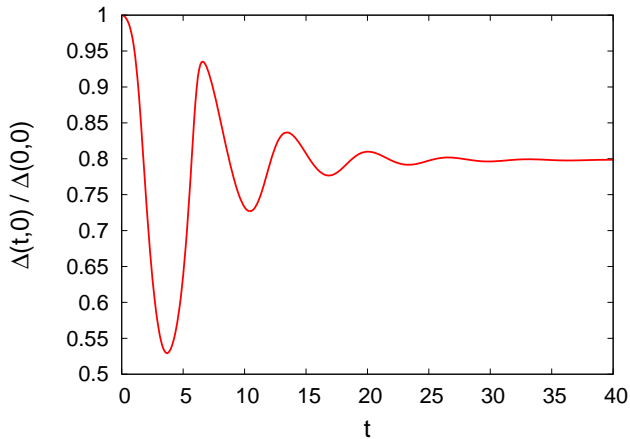}
\end{center}                                              
\caption{Quasiparticle gap (in units of the initial value) as a function of time.}
\label{fig:gap}                                                                       
\end{figure}

As the cascade progresses toward 
the ultraviolet, more 
oscillation modes of $\D(\rho)$ become 
excited, while the occupation number in each mode becomes smaller. 
Eventually, it becomes small enough
to require that the process is treated quantum-mechanically. We expect that, at that
point, the system will start to thermalize.
Thermalization will be slow, as the interaction between oscillation
modes with low
occupation numbers is weak, but the classical stage, shown in Fig.~\ref{fig:wave}, 
is quite fast,
with the characteristic timescale of order $R$. This is illustrated by Fig.~\ref{fig:gap},
which uses the same data as Fig.~\ref{fig:wave} and shows
the quasiparticle gap as a function of time. The gap is defined as the minimum of 
(\ref{eps}) with $\D(\rho)$ set to the instantaneous profile of the solution.
For the present dataset, the minimum is at $\rho =0$ at all times.
We see that the gap
rapidly relaxes to the final equilibrium value. The envelope 
is roughly exponential, with the decay constant (inverse relaxation time)
of about 0.2 (in units of $1/\Rini$). 
In applications to cold atoms, that would imply
a relaxation time of order $\xi / v_F$, where $\xi$ is the healing length of the superfluid.

To summarize, we have identified a mechanism by which paired Fermi
superfluids can relax to equilibrium even in
the absence of thermal quasiparticles or any additional degrees of freedom (such as
lattice phonons). The mechanism involves an ultraviolet cascade of the excess
energy and activates pairing amplitudes of fermions with various momenta,
including those far from the Fermi surface. It can operate entirely in the momentum
space, without requiring any coordinate dependence of the amplitudes. For a uniform quench,
the quasiparticle gap rapidly relaxes to the true equilibrium value.
Our results have been obtained for a specific strongly coupled 
superconductor by using string duality. The latter turns the distance 
from the Fermi surface into an extra coordinate, making
the cascade in the momentum space particularly easy to visualize.
The mechanism itself, however, seems completely generic. 
Indeed, we do not exclude that one
will see the same physics in the conventional mean-field theory, once
the pairing potential is given a suitable momentum dependence.

The author thanks M. Kruczenski for a discussion of the results.
This work was supported in
part by the U.S. Department of Energy grant \protect{DE-SC0007884}.


\begin{thebibliography}{99}
\bibitem{f1} The propagation speed of fermions in the normal state 
  has been set to one, $v_F = 1$, so lengths and times are in the same units.
\end{thebibliography}
\end{document}